\documentclass[preprint,aps,tightenlines,floatfix]{revtex4}
\newif\ifpdf
	\ifx\pdfoutput\undefined
	\pdffalse 
	\else
	\pdfoutput=1 
	\pdftrue
\fi

\ifpdf
	\usepackage[pdftex]{graphicx}
	\else
	\usepackage{graphicx}
\fi
\begin{document}
\ifpdf
	\DeclareGraphicsExtensions{.pdf, .jpg, .tif}
	\else
	\DeclareGraphicsExtensions{.eps, .jpg}
\fi
\title{Linear Optical CNOT Gate in the Coincidence Basis}
\author{ T. C. Ralph, N. K. Langford, T. B. Bell and A. G. White}
\affiliation{Centre for Quantum Computer Technology,
\\ Department of Physics,\\
University of Queensland, QLD 4072, Australia
\\Fax: +61 7 3365 1242  Telephone: +61 7 3365 3412 \\
email: ralph@physics.uq.edu.au\\}

\begin{center}
\scriptsize (16th December 2001)
\end{center}

\begin{abstract}
We describe the operation and tolerances of a non-deterministic, 
coincidence basis, quantum CNOT gate for photonic qubits.  It is 
constructed solely from linear optical elements and requires only a 
two-photon source for its demonstration.
\end{abstract}
\maketitle

\section{Introduction}

Qubits based on the polarisation state of individual photons have the 
advantage of low decoherence rates and are easily manipulated at the 
single qubit level.  Optical parametric amplification experiments 
have 
been very successful in producing and analysing a large range of two 
photon entangled states \cite{kwiat1,gisin,white1,white2}.  A key 
``trick'' in these types of experiments is to work in the coincidence 
basis in which only events where two photons are detected in the 
same, 
narrow time window are recorded.  The entangled state postselected 
this way may be a pure Bell state even though the total state is 
non-deterministic and may have experienced considerable mixing from 
photon loss.  Such systems are not scaleable in the quantum 
computational sense in their present form but none-the-less provide 
an 
excellent testing ground for quantum information concepts.  Useful 
application of this type of technology seems much closer in the realm 
of quantum communications.

A key two qubit gate is the Controlled Not (CNOT) gate.  A 
deterministic CNOT gate would require either very high 
non-linearities 
\cite{mil88} or very complex linear networks \cite{KLM}.  Building on 
the latter ideas a linear, coincidence basis CNOT has been described 
\cite{ral} which could be a useful test-bed.  However it requires a 
4-photon input, which is challenging.  Two photon, coincidence basis 
gates, which perform some, but not all of the operations of a CNOT 
gate have also been described \cite{bou,pan,fra}.

In this paper we discuss a linear, coincidence basis gate which 
performs all the operations of a CNOT gate and requires only a two 
photon input \cite{ralpat}.  In section 2 we describe its 
construction 
and ideal operation.  In section 3 we consider the effect of 
imperfections in its construction, particularly focusing on the 
effect 
of beamsplitter and mode-matching errors on the gates efficacy as a 
Bell state analyser.  In section 4 we conclude.  Recently Hofmann and 
Takeuchi have independently described a very similar gate 
\cite{hof}.  
Our analysis should also apply to their construction.

\section{The Gate}

The gate is shown in Fig.1.  All beamsplitters, $B1$, $B2$, $B3$, 
$B4$, and $B5$, are assumed asymmetric in phase.  That is, it is 
assumed that the operator input/output relations (the Heisenberg 
equations) between the two input mode operators ($a_{in}$ and 
$b_{in}$) and the corresponding output operators ($a_{out}$ and 
$b_{out}$) for the beamsplitters have the general form
\begin{eqnarray}
    a_{out}=\sqrt{\eta}a_{in}+\sqrt{1-\eta}b_{in}\nonumber\\
    b_{out}=\sqrt{1-\eta}a_{in}-\sqrt{\eta}b_{in}
\end{eqnarray}
where $\eta$ ($1-\eta$) is the reflectivity (transmittivity) of the 
beamsplitter.  Reflection off the bottom produces the sign change 
except for $B1$ and $B2$ which have a sign change by reflection off 
the top.  This phase convention simplifies the algebra but other 
phase 
relationships will work equally well in practice.  Beamsplitters $B3$ 
and $B4$ are both 50:50 ($\eta=1/2$).  The beamsplitters $B1$, $B2$ 
and $B5$ have equal reflectivities of one third ($\eta=1/3$).

We employ dual rail logic such that the ``control in'' qubit is 
represented by the two bosonic mode operators $c_{H}$ and $c_{V}$.  A 
single photon occupation of $c_{H}$ with $c_{V}$ in a vacuum state 
will be our logical 0, which we will write $|H \rangle_{c}$ (to avoid 
confusion with the vacuum state).  Whilst a single photon occupation 
of $c_{V}$ with $c_{H}$ in a vacuum state will be our logical 1, 
which 
we will write $|V \rangle_{c}$.  Superposition states can also be 
formed via beamsplitter interactions.  Similarly the ``target in'' is 
represented by the bosonic mode operators $t_{H}$ and $t_{V}$ and the 
states $|H \rangle_{t}$ and $|V \rangle_{t}$, with the same 
interpretations as for the control.  The use of $H$ and $V$ to 
describe the states of the qubits of course alludes to the usual 
encoding in polarisation \cite{encoding}.  To go from polarisation 
encoding to dual rail spatial encoding and vice versa in the lab 
requires a polarising beamsplitter and half-wave plate.

The Heisenberg equations relating the control ($c_{H},c_{V}$) and 
target ($t_{H},t_{V}$) input modes to the their corresponding outputs 
are
\begin{eqnarray}
  c_{H_O} & = & {{1}\over{\sqrt{3}}}(\sqrt 2  v_c+  c_H)\nonumber\\
  c_{V_O} & = & {{1}\over{\sqrt{3}}}(-  c_V+  t_H+  t_V)\nonumber\\
  t_{H_O} & = & {{1}\over{\sqrt{3}}}(  c_V+  t_H+   v_t)\nonumber\\
  t_{V_O} & = & {{1}\over{\sqrt{3}}}(  c_V+  t_V-  v_t)\nonumber\\
  v_{c_O} & = & {{1}\over{\sqrt{3}}}(-  v_c+\sqrt 2  c_H)\nonumber\\
  v_{t_O} & = & {{1}\over{\sqrt{3}}}(  t_H+  t_V -  v_t)
\label{eq:cnotbasic}
\end{eqnarray}
Ancillary, vacuum input modes, $v_{c}$ and $v_{t}$, complete the 
network.  The gate operates by causing a sign shift in the 
interferometer formed by the splitting and remixing of the target 
modes, conditional on the presence of a photon in the $c_{V}$ mode.  
Thus the target modes swap if the control is in the state $|V 
\rangle_{c}$ but do not if the control is in state $|H \rangle_{c}$.  
This is always true when a coincidence is measured between the 
control 
and target outputs (photons are detected at the same time).  However 
such coincidences only occur one ninth of the time, on average.  The 
other eight times out of nine either the target or the control or 
both 
do not contain a photon.  This can be seen explicitly by calculating 
the output state of the system in the Schr\"{o}dinger picture.  
Consider the general input state
\begin{eqnarray}
|\phi \rangle & = & (\alpha |HH \rangle+\beta |HV \rangle
+\gamma |VH \rangle+\delta |VV
 \rangle ) |00 \rangle \nonumber\\
 & = & (\alpha c_{H}^{\dagger} t_{H}^{\dagger}+
 \beta c_{H}^{\dagger} t_{V}^{\dagger}+\gamma c_{V}^{\dagger} 
t_{H}^{\dagger}
 +\delta c_{V}^{\dagger} t_{V}^{\dagger} ) |0000 \rangle |00 \rangle
 \label{input}
\end{eqnarray}
where the ordering in the kets is $|n_{cH} n_{cV} n_{tH} n_{tV} 
\rangle |n_{vc} n_{vt} \rangle $ with $n_{cH}=c_{H}^{\dagger}c_{H}$ 
etc and we use the short hand $|1010 \rangle =|HH \rangle$ etc where 
appropriate.  For a time symmetric linear network such as that in 
Fig.1, the output state can be directly obtained from the input 
state, 
Eq.\ref{input}, by substituting input operators for the output 
operators given by Eq.\ref{eq:cnotbasic}.  Thus we obtain
\begin{eqnarray}
|\phi \rangle_{out} & = & (\alpha \; c_{H_{O}}^{\dagger} 
t_{H_{O}}^{\dagger}+
 \beta \; c_{H_{O}}^{\dagger} t_{V_{O}}^{\dagger}+\gamma \;
c_{V_{O}}^{\dagger}
 t_{H_{O}}^{\dagger}
 +\delta \; c_{V_{O}}^{\dagger} t_{V_{O}}^{\dagger} ) |0000 \rangle 
|00
\rangle
 \nonumber\\
 & = & {{1}\over{3}} \{ \alpha |HH \rangle +\beta |HV \rangle +\gamma
 |VV \rangle +\delta |VH \rangle \nonumber\\
 & & +\sqrt{2}(\alpha+\beta) |0100 \rangle |10 \rangle+
 \sqrt{2}(\alpha-\beta) |0000 \rangle |11 \rangle+
 (\alpha+\beta) |1100 \rangle |00 \rangle \nonumber\\
& &  +(\alpha-\beta) |1000 \rangle |01 \rangle+
 \alpha |0010 \rangle |10 \rangle+\beta |0001 \rangle |10 \rangle
 \nonumber\\
& &  -(\gamma+\delta) |0200 \rangle |00 \rangle-
 (\gamma-\delta) |0100 \rangle |01 \rangle+
 \gamma |0020 \rangle |00 \rangle \nonumber\\
& &  +(\gamma-\delta) |0010 \rangle |01 \rangle+
 (\gamma+\delta) |0011 \rangle |00 \rangle
+(\gamma-\delta) |0001 \rangle |01 \rangle+
 \delta |0002 \rangle |00 \rangle \}
 \label{output}
\end{eqnarray}
The state postselected in the coincidence basis is then just
\begin{eqnarray}
|\phi \rangle_{cb} & = & \alpha |HH \rangle +\beta |HV \rangle +\gamma
 |VV \rangle +\delta |VH \rangle
 \label{cbout}
\end{eqnarray}
occuring with probability one ninth.  The relationship between 
Eq.\ref{input} and Eq.\ref{cbout} is a CNOT transformation.

It is also useful to look at the coincidence number expectation 
values, obtained directly from the Heisenberg equations 
(Eq.\ref{eq:cnotbasic}).  These can be interpreted as the predicted 
output coincident count rates normalized to the input pair rate.  An 
example is given in Table 1 which shows the count rates for logical 
basis inputs.  A more interesting case is to use the four Bell-states,
\begin{eqnarray}
    |\psi^{\pm} \rangle & = & {{1}\over{\sqrt{2}}}
(|H \rangle_{c} |H \rangle_{t}\pm|V \rangle_{c} |V
\rangle_{t})\nonumber\\
|\phi^{\pm} \rangle & = & {{1}\over{\sqrt{2}}}(|H \rangle_{c} |V
\rangle_{t} \pm
|V \rangle_{c} |H \rangle_{t})
\end{eqnarray}
as inputs and to detect the control in the superposition basis by 
mixing the control outputs on a 50:50 beamsplitter before detection:
\begin{eqnarray}
  c_{S_1} &=& {{1}\over{\sqrt{2}}}(  c_{H_O} +   c_{V_O})\nonumber\\
  c_{S_2} &=& {{1}\over{\sqrt{2}}}(  c_{H_O} -   c_{V_O})
\label{eq:supbasis}
\end{eqnarray}
In Table 2 the count rates for this arrangement are presented showing 
the ability to distinguish all four Bell states (albeit with non unit 
efficiency).  Such a Bell state analyser could have significant 
applications in quantum communications.  In the next section we will 
use this application as an example in order to investigate the effect 
of non-optimal parameters on the gate.

\section{Non-Optimal Operation}

The accuracy with which the gate operates will be determined by how 
closely the parameters of the constructed gate correspond to those of 
the idealized gate of the previous section.  We can identify three 
potential sources of error: incorrect beamsplitter ratios; non-unit 
mode matching and; timing errors.  One advantage of working in the 
coincidence basis is that losses and detector inefficiency can be 
ignored because they take the system out of the coincidence basis and 
thus their only effect is to reduce the count rate.

{\it Timing Errors}.  Correct gate operation depends on 
indistinguishability of the paths taken by the two photons through 
the 
network.  This means that they must arrive simultaneously at the 
central beamsplitter to an accuracy of a fraction of their coherence 
length.  Photon coherence length in down conversion experiments is 
generally determined by pre-detection frequency filtering and can be 
of order one hundred wave-lengths.  Locking path lengths on this 
scale 
should not be a major problem.

{\it Beamsplitter ratios}.  The effect of non-optimal beamsplitter 
ratios can be investigated by deriving the operator equations 
(Eq.\ref{eq:cnotbasic}) more generally, with arbitrary beamsplitter 
ratios.  For simplicity we assume that the beamsplitters all came 
from 
the same ``production-run'' such that any deviation from the optimal 
value is common.  That is, we might suppose that both the 50:50 
beamsplitters actually have a reflectivity of $\eta'$ whilst the 
three 
33:67 beamsplitters all actually have reflectivities $\eta$.  The 
Heisenberg equations are then
\begin{eqnarray}
c_{H_{O}} &=& \sqrt{\eta} c_{H} + \sqrt{1-\eta}
v_{c} \nonumber \\
c_{V_{O}} &=& - \sqrt{\eta} c_{V} +
\sqrt{(1-\eta)\eta' } t_{H} + \sqrt{(1-\eta)(1-\eta' )} t_{V}
\nonumber \\
t_{H_{O}} &=&  \sqrt{\eta}(1-2 \eta' )
t_{V} + 2
\sqrt{\eta(1-\eta' )\eta' } t_{H}+
\sqrt{\eta\eta' } c_{V} + \sqrt{(1-\eta)(1-\eta' )} v_{t}
\nonumber \\
t_{V_{O}} &=& 2
\sqrt{\eta(1-\eta' )\eta' } t_{V} + \sqrt{\eta}
(1-2\eta') t_{H} + \sqrt{(1-\eta)(1-\eta' )}
c_{V_{O}} - \sqrt{(1-\eta)\eta' } v_{t} \nonumber \\
v_{c_{O}} &=& - \sqrt{\eta} v_{c} + \sqrt{1-\eta}c_{H_{O}} \nonumber\\
v_{t_{O}} &=& \sqrt{(1-\eta)(1-\eta' )} t_{H}+
\sqrt{(1-\eta)\eta' } t_{V} - \sqrt{\eta} v_{t}
\label{generalCNOTopEqns}
\end{eqnarray}
In general the effect of varying the beamsplitter ratios is input 
state dependent.  However for small deviations from the optimum 
values 
Bell state analysis is approximately state independent and serves as 
a 
useful diagnostic \cite{note2}.  In Fig.2 we plot the error 
probability in distinguishing the Bell states as a function of $\eta$ 
and $\eta'$ in the region close to their optimum values.  The 
dependence of the error probability on $\eta'$ is mirror imaged 
between the $|\psi^{\pm} \rangle$ and the $|\phi^{\pm} \rangle$ Bell 
states.  However this dependence is negligible in the region close to 
$\eta'=1/2$.  The dependence on $\eta$ is more pronounced.  For an 
$\eta$ of $1/3 \pm 0.01$ (and $\eta'$ of $1/2\pm 0.05$) error rates 
of 
about $0.7 \%$ are predicted.  Such uncertainties are standard with 
current beamsplitter technology, and we conclude that errors below 
1.0 
\% are realistic.

{\it Mode matching errors}.  Mode matching in non-classical 
interference experiments is generally quite difficult and may be 
identified as a major contributor to non-unit visibility.  Given the 
key role of non-classical interference in the CNOT gate we may expect 
mode matching errors to be of some significance.

In order to model the mismatch of input modes at the central 
beamsplitter, ancillary modes $ v_1$, $ v_2$ and $ v_3$ (originally 
in 
the vacuum state) are introduced to interact with the propagating 
mismatch mode.  The additional output modes are labelled $ c_{V_m}$, 
$ 
c_{H_m}$ and $t_{V_m}$ (see Fig.3).  The mode $ c_v$ is assumed to be 
the source of the mismatch, after having passed through some kind of 
optical element that has misaligned it.
\begin{eqnarray*}
  c_{v_1} &=& \sqrt \xi \;  c_V +\sqrt{1-\xi} \;  v_1\\
  c_{v_2} &=& -\sqrt{1-\xi} \;  c_V +\sqrt \xi \;  v_1
\end{eqnarray*}
The parameter $\xi$ quantifies the degree of mode matching between 
the 
control and target modes at the central beamsplitter.  So long as the 
modes are matched reasonably well, $c_{v_{1}}$ can be considered a 
sort of ``primary'' mode.  It interacts with the output from 
beamsplitter $B_3$ in the same way as for the case neglecting mode 
matching.  The mismatch component $ c_{v_{2}}$ interacts only with 
the 
newly introduced vacuum modes.

The equations for the output modes of the quantum CNOT gate, 
including 
the effects of a mode mismatch, are
\begin{eqnarray}
  v_{c_O} & = & {{1}\over{\sqrt{3}}}(-  v_c+\sqrt 2  c_H)\nonumber\\
  c_{H_O} & = & {{1}\over{\sqrt{3}}}(\sqrt 2  v_c+  c_H)\nonumber\\
  c_{V_O} & = & {{1}\over{\sqrt{3}}}(-\sqrt \xi   c_V -\sqrt{1-\xi}
v_1 +  t_H+  t_V)\nonumber\\
  c_{V_m} & = & {{1}\over{\sqrt{3}}}(\sqrt{1-\xi} c_V-\sqrt \xi v_1 + 
\sqrt 2
v_2)\nonumber\\
  t_{H_O} & = & {{1}\over{\sqrt{3}}}(\sqrt\xi  c_V+  t_H+
\sqrt{1-\xi} v_1+  v_t)\nonumber\\
  t_{H_m} & = & {{1}\over{\sqrt{3}}}(-\sqrt{1-\xi}  c_V+ \sqrt \xi
v_1+ {{1}\over{\sqrt{2}}}   v_2 + \sqrt\frac{3}{2}   v_3) \nonumber\\
  t_{V_O} & = & {{1}\over{\sqrt{3}}}(\sqrt\xi
c_V +   t_V -\sqrt{1-\xi}   v_1-  v_t)\nonumber\\
  t_{V_m} & = & {{1}\over{\sqrt{3}}}(-\sqrt{1-\xi}  c_V + \sqrt\xi  
v_1
+{{1}\over{\sqrt{2}}}  v_2 - \sqrt\frac{3}{2}   v_3)\nonumber\\
  v_{t_O} & = & {{1}\over{\sqrt{3}}}(  t_H -   t_V -   v_t)
\label{eq:cnotmismatch}
\end{eqnarray}
Now, when measuring the coincidences, the detectors see a combination 
of the counts from both the primary modes and the mismatch modes (see 
Fig.3).  For example, when detecting coincidences of horizontally 
polarised photons, the count rate becomes
\begin{eqnarray}
\langle n_{cH_{D}} n_{tH_{D}} \rangle  &=&
\langle n_{cH_O}(n_{tH_O}+n_{tH_m}) \rangle \nonumber\\
&=& \langle n_{cH_{O}} n_{tH_{O}} \rangle
+ \langle n_{cH_{O}} n_{tH_{m}} \rangle  \textrm{ and similarly,} 
\nonumber\\
\langle n_{cH_{D}} n_{tV_{D}} \rangle  &=&
\langle n_{cH_{O}} n_{tV_{O}} \rangle  +
\langle n_{cH_{O}} n_{tH_{m}} \rangle  \nonumber\\
\langle n_{cV_{D}} n_{tH_{D}} \rangle  &=&
\langle n_{cV_{O}} n_{tH_{O}} \rangle  +
\langle n_{cV_{O}} n_{tH_{m}} \rangle  +
\langle n_{cV_{m}} n_{tH_{O}} \rangle +
\langle n_{cV_{m}} n_{tH_{m}} \rangle  \nonumber\\
\langle n_{cV_{D}} n_{tV_{D}} \rangle  &=&
\langle n_{cV_{O}} n_{tV_{O}} \rangle +
\langle n_{cV_{O}} n_{tV_{m}} \rangle +
\langle n_{cV_{m}} n_{tV_{O}} \rangle +
\langle n_{cV_{m}} n_{tV_{m}} \rangle
\label{eq:combmom}
\end{eqnarray}
These moments are summarised for logical inputs in Table 3.  As 
expected, the mode mismatch has not affected the CNOT operation when 
the control is ``off'' (i.e. when $c_{H}$ is occupied).  In this 
case, 
there is no interaction at beamsplitter $B_2$ (Fig.3) and thus no 
non-classical interference.  However, when the control is ``on'', the 
effects of the mismatch are noticeable.

Interestingly, the mismatch adds extra terms rather than 
redistributing the probabilities of the counts measured in the ideal 
case.  Coincidence events which previously were disallowed due to the 
non-classical interference can now appear as error events because of 
the mismatch.  Thus the probabilities that are being redistributed 
are 
those for the states that were not detected in the ideal case (the 
states which had been postselected out).

We now consider the performance of the gate as a Bell state analyser 
in the presence of mode mismatch.  As in the ideal case, another 
beamsplitter is added to the outputs of the control qubit.  Another 
ancillary mode $ v_4$ must be added to interact with the mismatch 
mode 
$ c_{V_m}$.

The beamsplitter outputs are given in the Heisenberg picture by
\begin{eqnarray}
  c_{S_{1O}} &=& {{1}\over{\sqrt{2}}}(  c_{H_O} +   
c_{V_O})\nonumber\\
  c_{S_{1M}} &=& {{1}\over{\sqrt{2}}}(  v_4 +   c_{V_m})\nonumber\\
  c_{S_{2O}} &=& {{1}\over{\sqrt{2}}}(  c_{H_O} -   
c_{V_O})\nonumber\\
  c_{S_{2M}} &=& {{1}\over{\sqrt{2}}}(  v_4 -   c_{V_m})
\label{eq:missupbasis}
\end{eqnarray}

Each detector receives counts from both of the modes incident on it, 
so the expectation values must be combined in a similar way to Eq.  
\ref{eq:combmom}.  The coincidence count rates are given in Table 4.  
Using $\xi=1$ yields the perfectly matched case calculated previously 
(see Table 2).  The error probability for Bell state discrimination 
is 
plotted in Fig.4.  For small mismatch the error is approximately 
equal 
to the percentage mismatch.  Clearly good Bell state discrimination 
will require accurate mode matching to the central beamsplitter.

\section{Conclusion}

We have described a non-deterministic quantum CNOT gate, that 
operates 
with one ninth efficiency, constructed solely from linear optical 
elements.  We have investigated the behaviour of the gate with 
variation in both the beamsplitter and mode match values and conclude 
that a demonstration is feasible with current optical technology.  
Aside from its value as a testbed system, such a gate could be made 
scaleable if photon-number QND detectors were added to each output.  
This latter system would also act as efficient Bell state analyser, 
which is an important component in some quantum algorithms, notably 
quantum teleportation.  

\begin{table}[!ht]
\centering
\begin{tabular}{|c|c|c|c|c|}
\hline
Input & $\langle n_{cH_{O}} n_{tH_{O}} \rangle $ &
$\langle n_{cH_{O}} n_{tV_{O}} \rangle$ &
$\langle n_{cV_{O}} n_{tH_{O}} \rangle$ &
$\langle n_{cV_{O}} n_{tV_{O}} \rangle$\\
\hline
\hline
$|H \rangle_{c} |H \rangle_{t}$ & $\frac{1}{9}$ & 0 & 0 & 0\\
\hline
$|H \rangle_{c} |V \rangle_{t}$ & 0 & $\frac{1}{9}$ & 0 & 0\\
\hline
$|V \rangle_{c} |H \rangle_{t}$ & 0 & 0 & 0 & $\frac{1}{9}$\\
\hline
$|V \rangle_{c} |V \rangle_{t}$ & 0 & 0 & $\frac{1}{9}$ & 0\\
\hline
\end{tabular}
\caption{Coincident expectation values calculated for the four 
logical 
basis inputs.}
\label{tab:baslog}
\end{table}
\begin{table}[ht!]
\centering
\begin{tabular}{|c|c|c|c|c|}
\hline
Input & $\langle n_{cS_{1}} n_{tH_{O}} \rangle $ &
$\langle n_{cS_{2}} n_{tV_{O}} \rangle$ &
$\langle n_{cS_{1}} n_{tH_{O}} \rangle$ &
$\langle n_{cS_{2}} n_{tV_{O}} \rangle$\\
\hline
\hline
$|\psi^{+} \rangle$ & $\frac{1}{9}$ & 0 & 0 & 0\\
\hline
$|\psi^{-} \rangle$ & 0 & $\frac{1}{9}$ & 0 & 0\\
\hline
$|\phi^{+} \rangle$ & 0 & 0 & $\frac{1}{9}$ & 0\\
\hline
$|\phi^{-} \rangle$ & 0 & 0 & 0 & $\frac{1}{9}$\\
\hline
\end{tabular}
\caption{Coincident expectation values calculated in the 
superposition 
basis for the four Bell states.}
\label{tab:basentsup2}
\end{table}
\begin{table}[!ht]
\centering
\begin{tabular}{|c|c|c|c|c|}
\hline
Input &  $\langle n_{cH_{D}} n_{tH_{D}} \rangle $ &
$\langle n_{cH_{D}} n_{tV_{D}} \rangle$ &
$\langle n_{cV_{D}} n_{tH_{D}} \rangle$ &
$\langle n_{cV_{D}} n_{tV_{D}} \rangle$\\
\hline
\hline
$|H \rangle_{c} |H \rangle_{t}$ & $\frac{1}{9}$ & 0 & 0 & 0\\
\hline
$|H \rangle_{c} |V \rangle_{t}$ & 0 & $\frac{1}{9}$ & 0 & 0\\
\hline
$|V \rangle_{c} |H \rangle_{t}$ & 0 & 0 & $\frac{2}{9}(1-\xi)$ &
$\frac{1}{9}$\\
\hline
$|V \rangle_{c} |V \rangle_{t}$ & 0 & 0 & $\frac{1}{9}$ &
$\frac{2}{9}(1-\xi)$\\
\hline
\end{tabular}
\caption{As for Table I, now allowing for mode matching $\xi$.  (For 
perfect mode match, $\xi=1$; for complete mode mismatch $\xi=0$).}
\label{tab:mislog}
\end{table}

\begin{table}[ht!]
\centering
\begin{tabular}{|c|c|c|c|c|}
\hline
Input & $\langle n_{cS_{1}} n_{tH_{D}} \rangle $ &
$\langle n_{cS_{2}} n_{tH_{D}} \rangle$ &
$\langle n_{cS_{1}} n_{tV_{D}} \rangle$ &
$\langle n_{cS_{2}} n_{tV_{D}} \rangle$\\
\hline
\hline
$|\psi^{+} \rangle$ &
$\frac{1}{18}(1+\sqrt\xi)$ & $\frac{1}{18}(1-\sqrt\xi)$ &
$\frac{1}{18}(1-\xi)$ & $\frac{1}{18}(1-\xi)$\\
\hline
$|\psi^{-} \rangle$ &
$\frac{1}{18}(1-\sqrt\xi)$ & $\frac{1}{18}(1+\sqrt\xi)$ &
$\frac{1}{18}(1-\xi)$ & $\frac{1}{18}(1-\xi)$\\
\hline
$|\phi^{+} \rangle$ &
$\frac{1}{18}(1-\xi)$ & $\frac{1}{18}(1-\xi)$ &
$\frac{1}{18}(1+\sqrt\xi)$ & $\frac{1}{18}(1-\sqrt\xi)$\\
\hline
$|\phi^{-} \rangle$ &
$\frac{1}{18}(1-\xi)$ & $\frac{1}{18}(1-\xi)$ &
$\frac{1}{18}(1-\sqrt\xi)$ & $\frac{1}{18}(1+\sqrt\xi)$\\
\hline
\end{tabular}
\caption{As for Table II, now allowing for mode match $\xi$.}
\label{tab:misentsup2}
\end{table}
\begin{figure}[ht!]
\centerline {\includegraphics[width=0.8 \columnwidth]{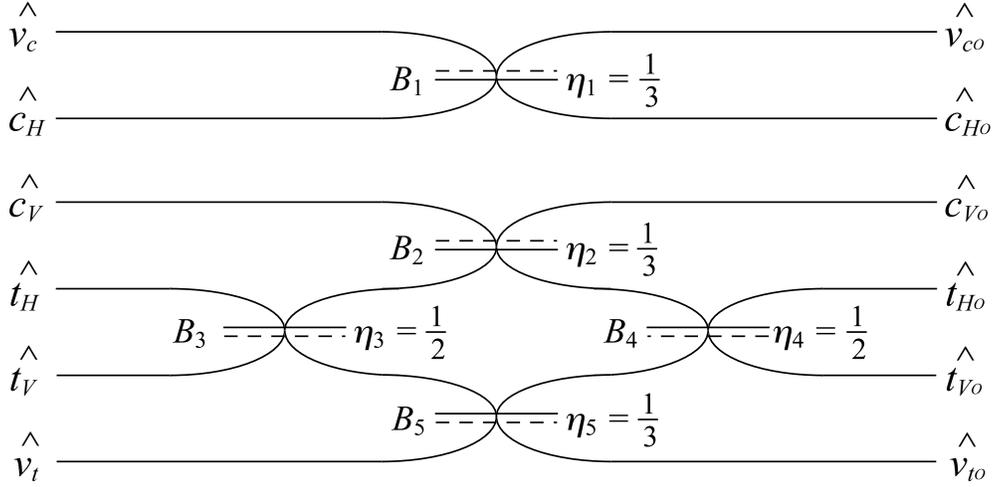}}
\footnotesize
\caption{Schematic of the coincidence CNOT gate. Dashing indicates 
the surface from which a sign change occurs upon reflection. The 
control modes are $c_{H}$ and $c_{V}$. the target modes are $t_{H}$ 
and $t_{V}$. The modes $v_{c}$ and $v_{t}$ are unoccupied ancillary 
modes.}
\end{figure}
\begin{figure}[ht!]
\centerline {\includegraphics[width=1.0 \columnwidth]{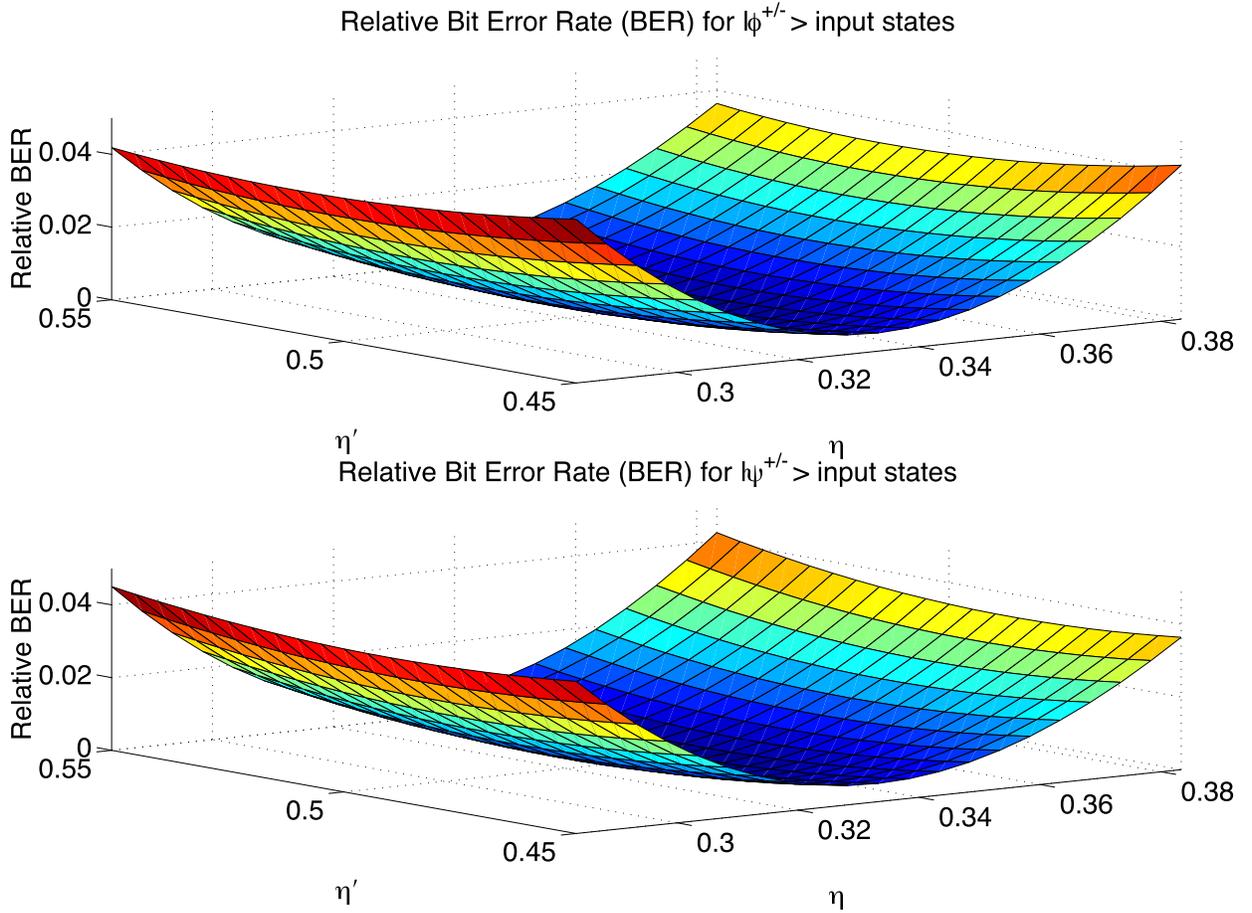}}
\footnotesize
\caption{Relative error rates, i.e. error rate/total rate, for Bell 
state analysis as a function of beamsplitter ratios close to the 
optimum values of $\eta'=1/2$ and $\eta=1/3$.}
\end{figure}
\begin{figure}[ht!]
\centerline {\includegraphics[width=0.8 \columnwidth]{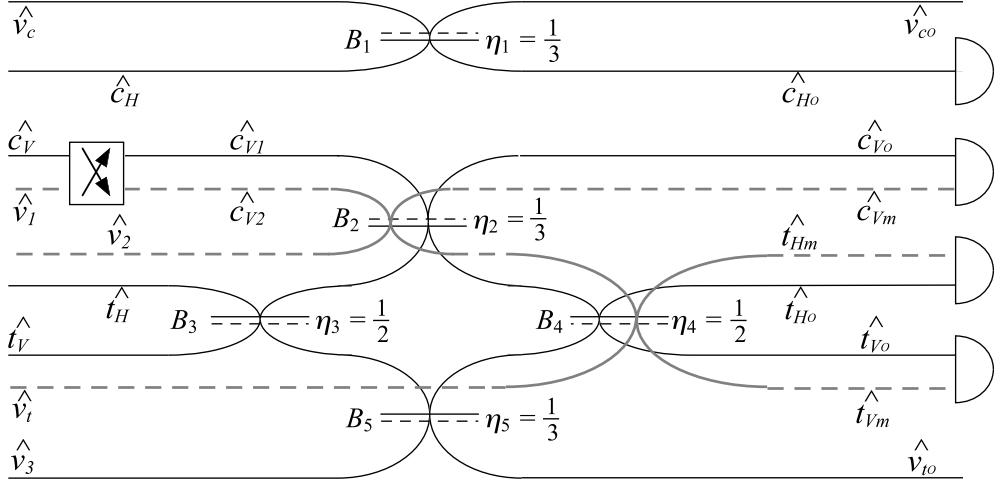}}
\footnotesize
\caption{Schematic diagram of the coincidence CNOT gate including the 
effects of mode matching. The mismatch is represented by splitting $  
c_V$ into two orthogonal modes $c_{V1}$ and $c_{V2}$. Ancillary modes 
$v_1$, $v_2$ and $v_3$ interact with the propagating mismatch.}
\end{figure}
\begin{figure}[ht!]
\centerline {\includegraphics[width=0.8 \columnwidth]{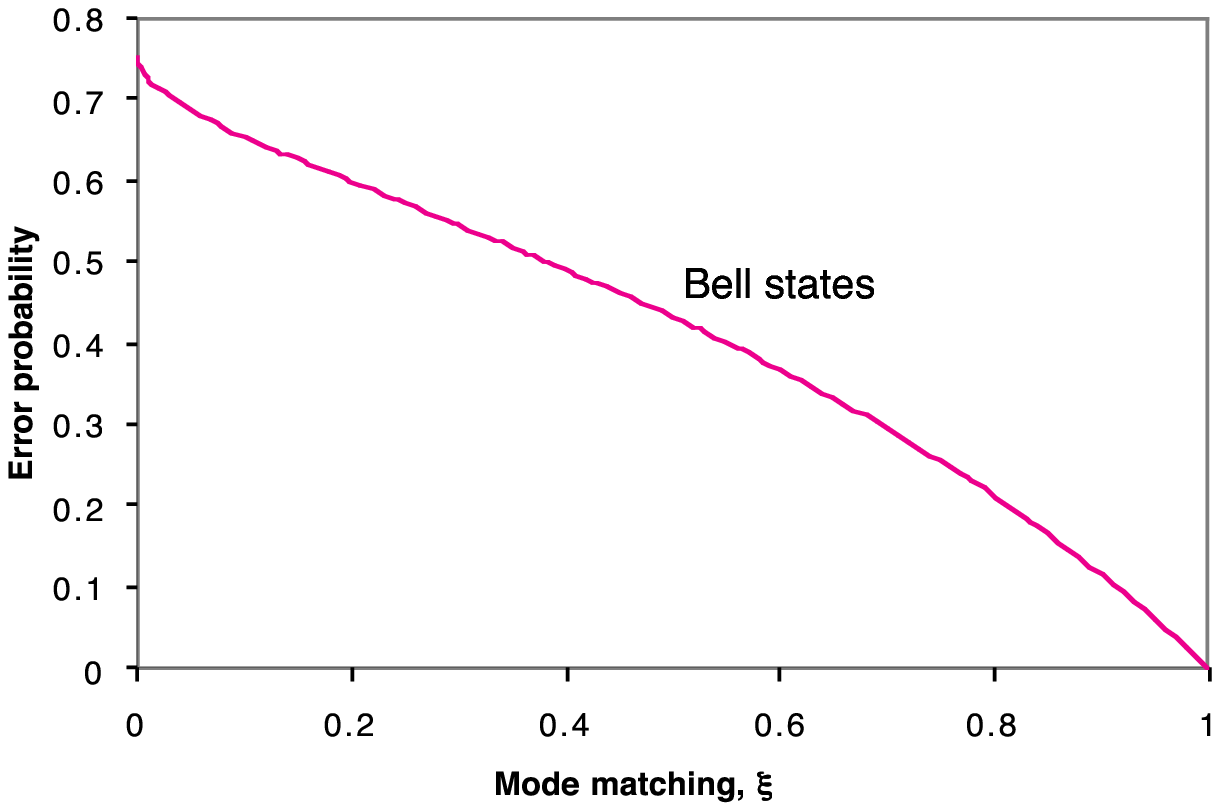}}
\footnotesize
\caption{Error probability as a function of mode matching for the 
four Bell states.}
\end{figure}


\begin{thebibliography}{99}
\bibitem{encoding} Of course, other degrees of freedom, such as 
orbital angular momentum, could be used in place of polarisation.
\bibitem{note2} Full characterization of the gate via quantum process 
tomography is currently under investigation.
\end{thebibliography}
\end{document}